\documentstyle[12pt,epsfig]{article}
\newcommand{\cm}{\mbox{\rm ~cm}}
\newcommand{\GeV}{\mbox{\rm ~GeV}}

\def\pb{\hbox{$\;\hbox{\rm pb}$}}

\newcommand{\lsim}{\raisebox{-0.5mm}{$\stackrel{<}{\scriptstyle{\sim}}$}}
\newcommand{\gsim}{\raisebox{-0.5mm}{$\stackrel{>}{\scriptstyle{\sim}}$}}
\newcommand{\Loooo}{$\sqrt{\lambda_{11}\lambda_{11}}$}
\newcommand{\Lootw}{$\sqrt{\lambda_{11}\lambda_{2j}}$}
\newcommand{\Looth}{$\sqrt{\lambda_{11}\lambda_{3j}}$}
\newcommand{\Lotot}{$\sqrt{\lambda_{12}\lambda_{12}}$}
\newcommand{\Lottw}{$\sqrt{\lambda_{12}\lambda_{2j}}$}
\newcommand{\Lotth}{$\sqrt{\lambda_{12}\lambda_{3j}}$}

\begin{document}
%
%
\begin{titlepage}
\vspace*{4.cm}
\begin{center}
\begin{Large}
\boldmath
\bf{A Search for Leptoquarks at HERA\\}
\unboldmath

\vspace*{2.cm}
H1 Collaboration \\
\end{Large}

\vspace*{1cm}

\end{center}

\vspace*{1cm}

\begin{abstract}
\noindent
A search for leptoquarks at HERA was performed in H1
using 1994 $e^+ p$ data corresponding to an integrated luminosity
of about $3 \pb^{-1}$.
Single leptoquarks were searched for in direct positron-quark fusion
processes taking into account possible decays into lepton-quark
pairs of either the first, the second, or the third generation.
No significant deviation from the Standard Model predictions
is found in the various final states studied and mass dependent
exclusion limits are derived on the Yukawa couplings of the
leptoquarks.
Compared with earlier results from an analysis of $e^-p$ data,
exclusion limits are considerably improved for leptoquarks
which could be produced via $e^+$-{\it valence~quark} fusion.
For leptoquarks with lepton flavour conserving couplings,
masses up to $275 \GeV$ (depending on the leptoquark type) are
excluded for coupling values larger than $\sqrt{ 4 \pi \alpha_{em}}$.
For leptoquarks with lepton flavour violating couplings,
masses up to $225 \GeV$ are excluded for the first time in a direct
search for couplings with leptons of the second or third generation
larger than $\sqrt{ 4 \pi \alpha_{em}}$.
Fourteen possible combinations of couplings are studied and
stringent exclusion limits comparable or better than any existing
direct or indirect limits are obtained for each leptoquark type.

\vspace{1cm}

\end{abstract}
\end{titlepage}

\vfill

%
\begin{flushleft}
%
 S.~Aid$^{14}$,                   
 V.~Andreev$^{26}$,               
 B.~Andrieu$^{29}$,               
 R.-D.~Appuhn$^{12}$,             
 M.~Arpagaus$^{37}$,              
 A.~Babaev$^{25}$,                
 J.~B\"ahr$^{36}$,                
 J.~B\'an$^{18}$,                 
 Y.~Ban$^{28}$,                   
 P.~Baranov$^{26}$
% ...
\end{flushleft}
\begin{flushleft} {\it
% ...
$ ^{12}$ DESY, Hamburg, Germany$^a$ \\
% ...
$ ^{14}$ II. Institut f\"ur Experimentalphysik, Universit\"at Hamburg,
                                     Hamburg, Germany$^ a$  \\
% ...
$ ^{18}$ Institute of Experimental Physics, Slovak Academy of
                Sciences, Ko\v{s}ice, Slovak Republic$^ f$\\
% ...
$ ^{25}$ Institute for Theoretical and Experimental Physics,
                                                 Moscow, Russia \\
 $ ^{26}$ Lebedev Physical Institute, Moscow, Russia$^ f$ \\
% ...
$ ^{28}$ LAL, Universit\'{e} de Paris-Sud, IN2P3-CNRS,
                            Orsay, France\\
 $ ^{29}$ LPNHE, Ecole Polytechnique, IN2P3-CNRS,
                             Palaiseau, France \\
% ...
$ ^{36}$ DESY, Institut f\"ur Hochenergiephysik,
                              Zeuthen, Germany$^ a$\\
 $ ^{37}$ Institut f\"ur Teilchenphysik,
          ETH, Z\"urich, Switzerland$^ i$\\
% ...
$ ^a$ Supported by the Bundesministerium f\"ur
        Forschung und Technologie, FRG
        under contract numbers 6AC17P, 6AC47P, 6DO57I, 6HH17P, 6HH27I,
        6HD17I, 6HD27I, 6KI17P, 6MP17I, and 6WT87P \\
% ...
$ ^f$ Supported by the Deutsche Forschungsgemeinschaft\\
% ...
$ ^i$ Supported by the Swiss National Science Foundation\\
  }
\end{flushleft}

\newpage
\section{Introduction}
\label{sec:intro}

The $ep$ collider HERA is particularly suited for the search for
leptoquark colour triplet bosons.
Such particles appear naturally in various unifying theories beyond the
Standard Model (SM) such as Grand Unified Theories and Superstring
inspired $E_6$ models, and in some Compositeness and Technicolour
models.
They could be produced singly as $s$-channel resonances at HERA
by the fusion of the $27.5 \GeV$ initial state lepton with a quark of
the $820 \GeV$ incoming proton.

In this paper we present a direct search for leptoquark resonances.
The analysis combines the H1 1993 $e^- p$ data~\cite{H1LQ94} and 1994
$e^+ p$ data which correspond respectively to an integrated
luminosity of $0.43 \pb^{-1}$ and $2.83 \pb^{-1}$.
Earlier direct searches at HERA for leptoquarks coupling to first
generation fermions were presented in~\cite{H1LQ94,H1LQ93,ZEUS93}.
The search for leptoquarks coupling also to either second or third
generation fermions is presented here for the first time.

\section{Phenomenology}
\label{sec:pheno}

We consider all possible scalar ($S_I$) and vector ($V_I$) leptoquarks
of weak isospin {\it I} with dimensionless couplings
$\lambda^{L,R}_{ij}$ to lepton-quark pairs, where {\it i} and {\it j}
indices denote lepton and quark generations respectively and
{\it L} or {\it R} is the chirality of the lepton.
Following the phenomenological ansatz of ref.~\cite{BUCH1987} which
introduces a general effective Lagrangian obeying the symmetries of the
SM, there are 10 different leptoquark isospin multiplets, with couplings
to left or right handed fermions
\footnote{A more detailed discussion of the leptoquark classification
          scheme may be found in a previous
          publication~\cite{H1LQ93} or
          in~\cite{BUCH1987,HERAWS}. }.
The search can be restricted to pure chiral couplings of the
leptoquarks given that deviations from lepton
universality in helicity suppressed pseudoscalar meson
decays have not been observed~\cite{DAVIDSON,LEURER}.
This restriction to couplings with either left- ($\lambda^L$) or
right-handed ($\lambda^R$) leptons
(i.e. $ \lambda^L \cdot \lambda^R \sim 0 $),
affects only two scalar leptoquarks ($S_0$ and $S_{1/2}$) and two
vector leptoquarks ($V_0$ and $V_{1/2}$).

We otherwise impose a minimal set of simplifying assumptions:
\begin{itemize}
  \item one of the leptoquark multiplets is produced dominantly;
  \item states in the leptoquark isospin doublets and triplets
        are degenerate in mass;
  \item there exists only one sizeable coupling of the leptoquarks
        to any given lepton generation.
\end{itemize}
This last assumption implies that we consider in the $s$-channel at HERA
only the production via $\lambda_{1j}$ where $j=1,2$
(implying a production cross-section scaling approximately in
$\lambda^2_{1j}$) followed by a decay either via the same coupling or
via a coupling $\lambda_{kl}$ where $k \neq 1$ and $l=1,2$ or $3$.
The former case, where we have a unique sizeable coupling to a single
lepton and to a single quark generation, corresponds to a ``diagonality''
requirement in the lepton and quark sector which circumvents the
stringent bounds coming from the absence of flavour changing neutral
current processes~\cite{DAVIDSON,LEURER}.
In the latter case, which admits more than one coupling of comparable
strengths, the above assumption guarantees that each coupling leads to
a distinct observable final state.
This is the case for instance for the striking lepton flavour
violating sub-processes $ e + q \rightarrow \mu + q' $ or $\tau + q'$.

In the $s$-channel, a leptoquark is produced at fixed mass $M=\sqrt{s x}$
where $\sqrt{s} \sim 300 \GeV$ is the energy available in the $ep$
centre-of-mass frame and $x$ is the incoming quark momentum fraction.
The resonance has an intrinsic width $\Gamma = \lambda^2 M/16\pi$ for
scalar and $\Gamma = \lambda^2 M/24\pi$ for vector leptoquarks.
Hence the production cross-section depends on the quark momentum density
in the proton and approximately scales with $\lambda^2$.
When involving first generation leptons, the decay of the leptoquark
into a lepton and a jet leads to signatures which are practically
indistinguishable event-by-event from SM neutral (NC) and charged
current (CC) deep inelastic scattering (DIS).
Statistically, the new signal may however be discriminated on the one
hand by the presence of a~peak in the invariant mass distribution
and, on the other hand, by the specific angular distribution of the
decay products which depends on the spin of the leptoquark.
For a vector leptoquark $d\sigma\,/\,dy \sim (1-y)^2$ where
$y=\frac{1}{2}\left(1+\cos{\theta^*}\right)$ is
the Bjorken scattering variable in DIS and $\theta^*$ is the decay
polar angle in the leptoquark centre of mass (CM) frame.
Scalars decay isotropically in their CM frame leading to a
constant $d\sigma\,/\,dy \;$.
These are markedly different from the $d\sigma\,/\,dy\sim\,y^{-2}$
distribution expected at fixed $x$ for the dominant $t$-channel
photon exchange in neutral current DIS events
\footnote{At high momentum transfer, $Z^0$ and $W$ exchanges become
          more important and contribute to less pronounced differences
          between signal and background.}.
Hence, first generation leptoquarks are searched for as DIS--like
events at high mass and high $y$.
In contrast, the  $ e + q \rightarrow \mu + q'$ or $\tau + q'$
(followed by a leptonic decay of the $\tau$) processes lead to exotic
signatures and are expected to be essentially background free
at high $Q^2$ ($Q^2 = s x y $).

\section{The H1 detector}
\label{sec:h1det}

A detailed description of the H1 detector can be found
elsewhere~\cite{H1DETECT}.
Here we describe only the components relevant for the present analysis
in which the event final state involves a lepton (positron, muon or
neutrino) with high transverse energy $E_{T,l}$, balanced
more-or-less by a large amount of hadronic $E_{T,h}$ flow.

The energy flow is measured in a finely segmented liquid
argon (LAr) sampling calorimeter~\cite{H1LARCAL} covering the polar
angle\footnote{The incoming proton moves by definition in the forward
          (i.e. $z > 0$) direction with $\theta=0^{\circ}$ polar
          angle.}
range 4$^{\circ} \le \theta \le$ 153$^{\circ}$ and all azimuthal
angles.
It consists of a lead/argon electromagnetic section followed by a
stainless-steel/argon hadronic section.
Electron energies are measured with a resolution
of $\sigma(E_e)/E_e \simeq$ $12$ \%/$\sqrt{E_e}\oplus1\% $ and hadron
energies with $\sigma(E_h)/E_h \simeq$ $50$ \%/$\sqrt{E_h}\oplus2\% \,$
{}~\cite{H1CALRES} ($E$ in \GeV) after software energy weighting.
The absolute energy scales are known to 2\% and 5\% for
electrons and hadrons respectively.
The angular resolution on the scattered electron measured from the
electromagnetic shower in the calorimeter is $\lsim \; 4 $ mrad.
A lead/scintillator electromagnetic backward calorimeter extends the
coverage at larger angles
(155$^{\circ} \le \theta \le$ 176$^{\circ}$).

Located inside the calorimeters is the tracking system used here to
determine the interaction vertex and to measure the track associated
to the final state lepton in exotic event topologies.
The main components of this system are central drift and proportional
chambers (25$^{\circ} \le \theta \le$ 155$^{\circ}$), a forward track
detector  (7$^{\circ} \le \theta \le$ 25$^{\circ}$) and backward
proportional chambers (155$^{\circ} \le \theta \le$ 175$^{\circ}$).
The tracking chambers and calorimeters are surrounded
by a superconducting solenoid coil providing a uniform field of
$1.15${\hbox{$\;\hbox{\rm T}$}}
within the tracking volume.
The instrumented iron return yoke surrounding this coil is used here to
measure leakage of hadronic showers.
The luminosity is determined from the rate of the Bethe-Heitler process
$e p \rightarrow e p \gamma$ measured in a luminosity monitor
as described in~\cite{H1DETECT}.

\section{Analysis}
\label{sec:data7s}

\subsection{DIS-like signatures}
\label{sec:dislike}

For leptoquarks decaying via a coupling to a positron or a neutrino,
we are searching for DIS-like signatures.

\begin{flushleft}
{\bf $e+q$ final states}
\end{flushleft}
\noindent

The selection of event candidates for $e+q$ final states relies
essentially on electron finding and energy-momentum conservation
cuts. We require:
\begin{enumerate}

  \item a primary interaction vertex in the range
        $\mid z - \bar{z} \mid < 35 \cm$ with $\bar{z} = 3.4 \cm$;

  \item an electron with $E_{T,e} = E_e \sin \theta_e > 7 \GeV$ and
        $10^{\circ} \le \theta_e \le  145^{\circ}$ (a range covered
        by the H1 LAr calorimeter).
        There must be either only one electron candidate or
        the candidate with highest $E_{T,e}$ must be at largest
        polar angle;

  \item a total missing transverse momentum
        $P_{T,miss} \approx
        \sqrt{\left(\sum E_x \right)^2 + \left(\sum E_y \right)^2}
        \leq 15 \GeV$  summed over all energy depositions $i$ in the
        calorimeters, with $ E_x^i = E^i \sin \theta^i \cos \phi^i $
        and $ E_y^i = E^i \sin \theta^i \sin \phi^i $;

  \item a minimal loss of longitudinal momentum in the direction
        of the incident electron, i.e.
        $ 43 \leq \sum \left(E - P_z\right) \leq 63 \GeV$, with
        $ P_z^i \approx E^i \cos \theta^i $;

  \item less than $5 \GeV$ in total in the backward calorimeters
        ($\theta \; \gsim \; 152^{\circ}$);

  \item a Bjorken $y_e$, measured from the final state electron,
        satisfying $ 0.05 < y_e < 0.95 $;

  \item when two electron candidates are found, these two candidates
        must not be balanced in $E_{T,e}$ and in azimuth,
        i.e. $E^1_{T,e} / E^2_{T,e} > 1.25$ and
        $ \mid \Delta \phi_{1,2} - 180^{\circ} \mid > 2^{\circ} $;

  \item the event must survive a set of halo and cosmic muon filters;

  \item the event must be accepted by LAr electron or transverse
        energy triggers~\cite{H1DETECT} and be properly in time
        relative to interacting bunch crossings.

\end{enumerate}
The electron identification, which relies on electromagnetic shower
estimators and an isolation criteria, was described in~\cite{H1LQ94}.
Cut (1) mainly suppresses beam--wall, beam--residual gas and, with (8),
background from cosmic rays and halo muons.
Cuts (2) and (4) provide a powerful rejection of photoproduction
contamination and, with (3), eliminate DIS charged current events.
Cut (4) also suppresses DIS NC-like events
with a very hard $\gamma$ emitted from the initial state
electron. Cut (5) further suppresses DIS (or photoproduction)
events at small momentum transfer with a misidentified electron
in the LAr calorimeter.
Cut (6) avoids the high $y_e$ region where the largest radiative
corrections are expected and the low $y_e$ region where the $y_e$
and $x_e$ resolutions deteriorate severely.
Cut (7) removes QED Compton events.

In total, 1800 NC-like events satisfy all the above requirements in the
mass range $ M_e > 45 \GeV$, where $M_e$ is reconstructed
from the final state electron energy and polar angle using
$$ M_e = \sqrt{ s x_e } = \sqrt{\frac{Q^2_e}{y_e}}, \;\;\;\;
     Q^2_e = \frac{E^2_{T,e}}{1-y_e}, \;\;\;\;
       y_e = 1 - \frac{E_e - P_{z,e}}{2E_e^0} \;\;\;\;  $$
where $E_e^0$ is the incident electron beam energy.

In order to compare with SM expectations for DIS NC, we make use
of the LEPTO Monte Carlo event generator~\cite{INGELMAN}, which includes
the lowest order electroweak scattering process with QCD corrections to
first order in $\alpha_s$, complemented by leading-log parton showers
and hadronisation~\cite{PYTHIA}.
The parton densities in the proton used throughout are taken from the
MRSH~\cite{MRSHSF} parametrisation which is close to recent $F_2$
structure function measurements at HERA (see~\cite{HERASF}).
All generated events (corresponding to $\sim 2.5$ times the integrated
luminosity for the $e^+p$ data) were passed through a detailed
Monte Carlo simulation of the H1 detector followed by the full analysis
chain.
We estimate a mean expected NC background of 1680 events with a
systematic uncertainty of $\pm$ 102 events.
The systematic uncertainty on the expected background comes from
the luminosity measurement (1.8\%), from the absolute energy
calibration (which translates to a 10\% effect) and from
finite Monte Carlo statistics (1.5\%).
Taking into account this systematic error and Poisson statistics, this is
compatible to less than a standard deviation with the 1800 events
observed in our measurement.
At largest masses, $M_e > 150 \GeV$, $43$  events are observed in
excellent agreement with the SM mean of $37.2 \pm 8.2$ (syst.) expected.

We further investigated possible contamination from direct and
resolved photoproduction events of light and heavy quarks.
This was studied on the one hand by looking for events with an electron
tagged in the H1 luminosity detector when removing cut (5).
This detector has an acceptance of $\sim 15\%$ for photoproduction
events with comparable total transverse energy.
No such events are found in the data.
On the other hand, an upper limit on the contamination was
derived from Monte Carlo studies based on the PYTHIA
generator~\cite{PYTHIA}.
The parton densities in the photon are taken from the
GRV-G LO parametrisation~\cite{SFGRVGLO}.
We estimate the contamination to be $< 0.7 \%$ for
$\gamma + p \rightarrow jet + jet$ (direct and resolved photon
processes) and  $< 0.6\%$ from heavy flavour pair production by
boson--gluon fusion processes.

%
\begin{figure}[htb]
  \begin{center}
    \begin{tabular}{cc}
      \mbox{\epsfxsize=7.5cm \epsffile{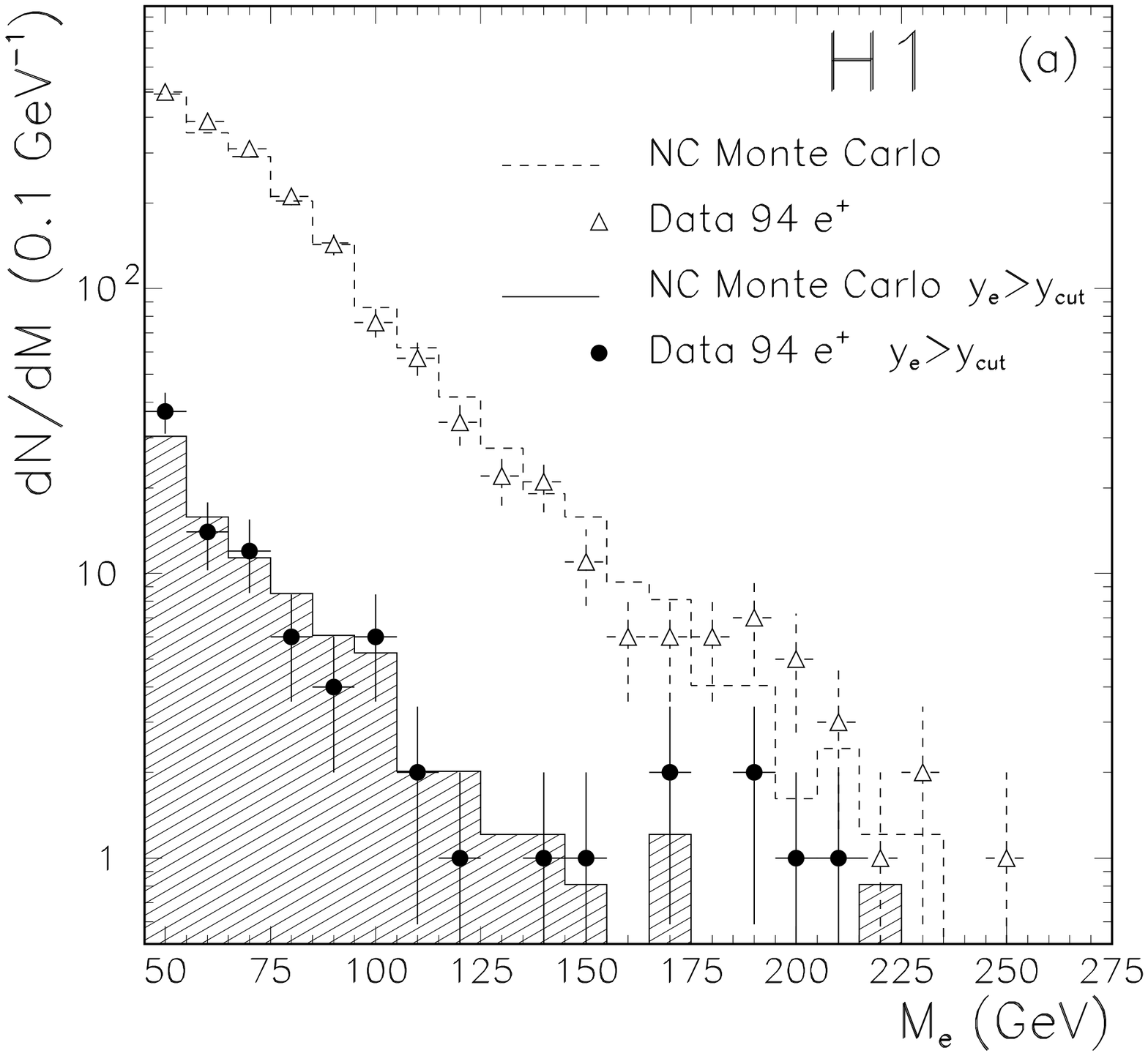}}
      \mbox{\epsfxsize=7.5cm \epsffile{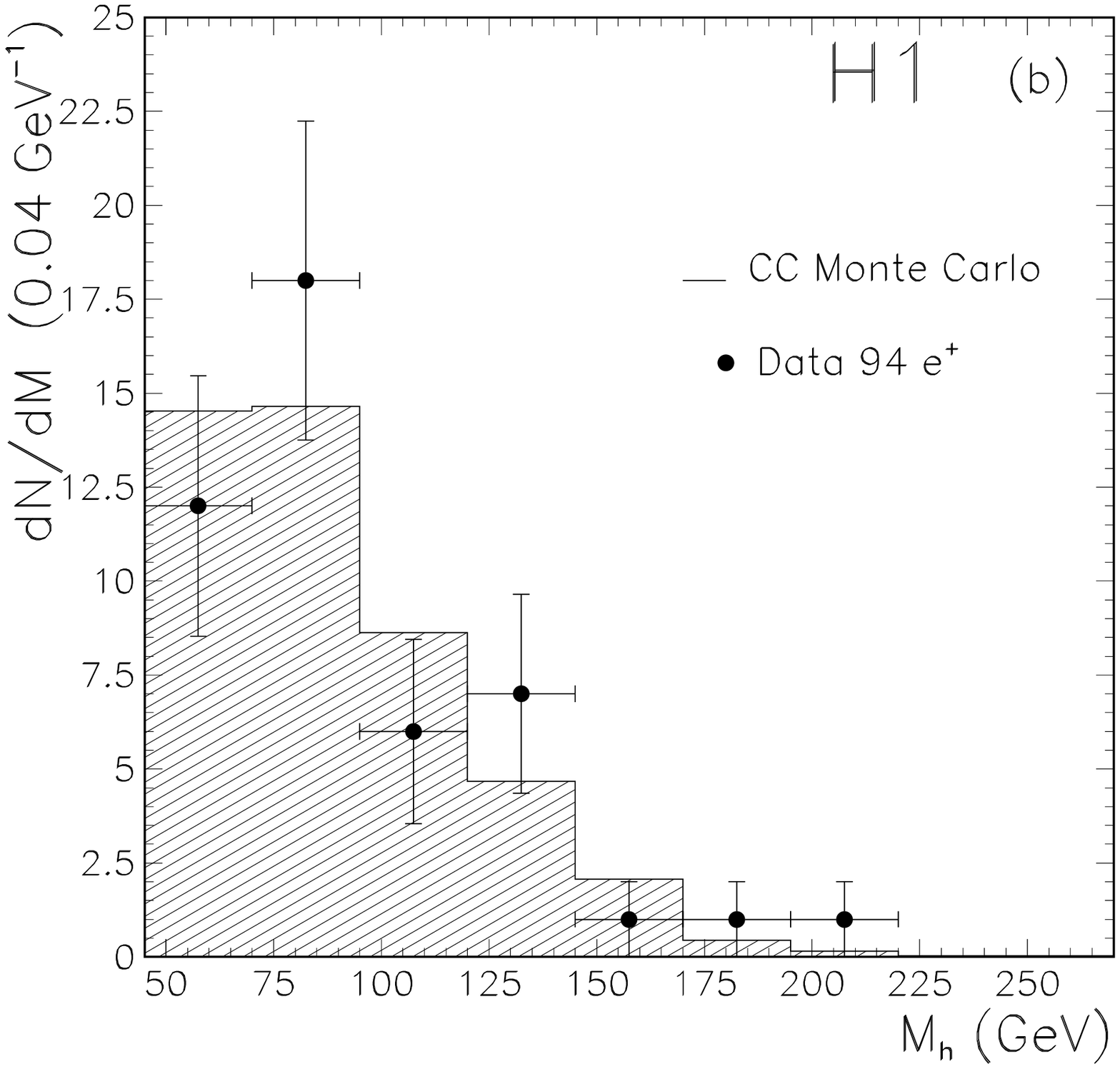}}
    \end{tabular}
  \end{center}
\caption[~]
{\small  Mass spectra for $e+q$ (a) and $\nu+q$ (b) final states
         for data (closed points) and DIS Monte Carlo (shaded
         histograms).
         Also shown are the data (open triangles) and DIS NC
         Monte Carlo (dashed histogram) compared before the final $y_e$
         cut (see text). }
 \label{fig:dndmlq}
\end{figure}

The comparison of the measured data and DIS NC predictions
is shown in the mass spectra of Fig.~\ref{fig:dndmlq}a.
The measured mass spectrum is seen before and after having applied a
mass dependent $y_e$ cut which ranges from $0.5$ around $M_e = 45 \GeV$
to $0.10$ around $M_e = 250 \GeV$.
This cut is designed~\cite{H1LQ94} via Monte Carlo studies
to optimize the signal significance for scalar leptoquark searches,
given the expected background.
A similar but less severe $y_e$ cut is optimized separately for vector
leptoquark searches.

In both cases the SM model expectations are in excellent agreement with
the data.
After the $y_e$ cut, we observe 91 events while $84.0 \pm 11.7$ (syst.)
events are expected for $e + q$ data at $M_e > 45 \GeV$.
At large masses, $M_e > 100 \GeV$, $13$ events are observed in excellent
agreement with the mean SM expectation of $12.4 \pm  3.8$ (syst.).
At largest masses, $M_e > 150 \GeV$, 7 events are observed in agreement
with the mean expectation of $3.5 \pm 2.0$ (syst.).

\begin{flushleft}
{\bf $\nu + q$ final states}
\end{flushleft}
\noindent
For $\nu+q$ final states the selection mainly relies
on the absence of an electron candidate and large missing momenta.
We impose the same primary vertex cuts and apply the same halo and
cosmic muon filters as for the $e+q$ analysis.
In addition, we require for the event candidates:

\begin{enumerate}

  \item no electron satisfying the requirements of the $e + q$ selection;

  \item $P_{T,miss} > 15 \GeV$;

  \item a total transverse energy
        $E_T \approx \sum \mid \vec{P}_T \mid$ roughly
        matching the total missing transverse momentum
        $P_{T,miss}$ such that $(E_T - P_{T,miss})/E_T < 0.5$;

  \item at least three charged tracks linked to the primary vertex
        when the polar angle $\theta_h$ associated with the hadronic
        transverse momentum flow
        ($ \theta_h \approx \tan^{-1} P_{T,miss}/E_T$)
        is within the angular range
        $35^{\circ} \le \theta_h \le  155^{\circ}$;

  \item the event must be accepted by the LAr missing transverse energy
        trigger~\cite{H1DETECT} and be properly in time relative
        to interacting bunch crossings.

\end{enumerate}
Cuts (1), (2) and (3) provide a powerful rejection
of photoproduction contamination and eliminate DIS NC events.
Halo and cosmic muons which are not accompanied by energy flow
in the forward region are rejected by cut (4).
We are left at this point with 252 event candidates.
We then apply more stringent cosmic and halo filtering algorithms which
are cross-checked by visual scan.

In total, 46 CC-like events satisfy all above requirements in the
relevant mass range at $ M_h > 45 \GeV$ where $M_h$ is reconstructed
by summing over all visible final state ``hadrons'' using
$$ M_h = \sqrt{\frac{Q^2_h}{y_h}}, \;\;\;\;
     Q^2_h= \frac{P^2_{T,miss}}{1-y_h},\;\;\;\;
       y_h=\frac{\sum \left(E-P_z\right)}{2E_e^0}.\;\;\;\; $$
This is in excellent agreement with the SM expectation of
43.2 $\pm$ 6.8 (syst.) events obtained using the Monte Carlo DJANGO
generator~\cite{DJANGO} which includes all radiative
channels and QCD dipole parton showers.
The LAr trigger efficiency losses~\cite{CCPAPER} were folded in.
The systematic errors coming from luminosity and energy calibration
are similar as for the $e+q$ channel but here the contribution of the
finite Monte Carlo statistic is negligible.
The mass spectra for the measured data and DIS CC Monte Carlo
predictions shown in Fig.~\ref{fig:dndmlq}b are in excellent
agreement over the full mass range.

\subsection{Exotic signatures}
\label{sec:exotic}

\begin{flushleft}
{\bf $\mu + q$ final states}
\end{flushleft}
\noindent
For leptoquarks coupling to a second generation lepton,	 leading to
$\mu+q$ final states, we can make use of the
cuts for the $\nu + q$ event sample
since such events are characterized by a large $P_{T,miss}$
when calculated from calorimetric quantities.
Such $P_{T,miss}$ is only slightly affected by the minimal energy
deposition of the muon.
In addition, we require:
\begin{enumerate}

  \item a charged track linked to the primary vertex within
         $\Delta \theta \, , \, \Delta \phi < 15^{\circ}$
        around the ``muon'' angle deduced from the kinematical
        constraints using the hadronic flow.

\end{enumerate}
The efficiency to find such a charged track was derived from
the $e+q$ event sample by searching on the basis of the hadronic energy
flow for the electron track associated with the electron candidate.
This efficiency is found to be $\sim 90\%$ for masses in the range
$45 \le M_h \le 225 \GeV$.
Higher masses lead to a charged lepton falling in an angular
domain where no such empirical determination of the efficiency
could be made.
Cut (1) offers powerful rejection against CC events as it imposes
a charged track opposite in azimuth to the current jet.
We require no explicit tagging of this ``muon'' track candidate in the
instrumented iron.

We observe no candidate satisfying the $\mu+q$
selection\footnote{It
          should be noted that the $e^+ p \rightarrow \mu^+ X$ event
          observed by H1 and discussed in~\cite{H1MUEV} fails
          significantly the kinematical constraints required of
          a $ e q \rightarrow  LQ \rightarrow  \mu q' $ event.}.

\begin{flushleft}
{\bf $\tau+q$ final states}
\end{flushleft}
\noindent
For $\tau + q$ final states, the analysis is restricted to leptonic
decays of the $\tau$
( $\tau^+ \rightarrow \mu^+ \nu_{\mu} \bar{\nu_{\tau}}$ and
  $\tau^+ \rightarrow e^+ \nu_e \bar{\nu_{\tau}}$ )
which constitute $\sim 36\%$ of all $\tau$ decays.
In the range of large leptoquark masses considered here, the $\tau$
decay products are generally strongly boosted in the $\tau$ direction.
Hence, the search strategy consists of looking either for $e+q$ or
$\mu + q$ final states where the $e$ or the $\mu$ angle is centered
on the $\tau$ angle deduced from the kinematical constraints
using the hadronic energy flow.

For the $\tau^+ \rightarrow \mu^+ \nu_{\mu} \bar{\nu_{\tau}}$ channel,
we can make use of the analysis cuts of the above $\mu + q$
event sample. The $P_{T,miss}$ measured in the calorimeters
is only slightly reduced on average on the quark side when the
leptoquark coupling involves a heavy quark which can undergo
a semi-leptonic decay. No candidates satisfy these cuts.
For the $\tau^+ \rightarrow e^+ \nu_e \bar{\nu_{\tau}}$ channel we
apply a set of cuts identical to those for the $\nu + q$ selection
except for the following:
\begin{enumerate}
  \item an electron satisfying the requirements of the $e + q$ selection,
        but found within
        $\Delta \theta \, , \, \Delta \phi < 15^{\circ}$
        around the predicted
        $\tau$ angle deduced from the kinematical constraints
        using the hadronic energy flow;
  \item $P_{T,miss} > 10 \GeV$ pointing in the azimuthal direction of
        the $\tau$.
\end{enumerate}
Cut (1) imposes a NC-like topology to events otherwise satisfying
the $\nu + q$ selection, hence suppressing DIS CC events.
The $P_{T,miss}$ constraint of cut (2) suppress DIS NC events.
We observe no candidate satisfying these cuts for $ M_h \ge 45 \GeV$.

\section{Results}
\label{sec:results}

In the absence of any significant deviation from the SM expectations,
we now derive rejection limits for Yukawa couplings as a function
of mass.
For each contributing channel, we use the number of
observed events, the signal detection efficiencies and the
expected number of background events within a mass bin of
variable width (adapted to the expected mass resolution)
which slides over the accessible mass range.
The statistical procedure which folds in channel per channel the
statistical and systematic errors is described in detail
in~\cite{H1LQ94}.
The signal detection efficiencies are typically determined over a
coupling-mass grid with steps in mass of $25 \GeV$ and for
coupling values corresponding roughly to the expected ultimate
sensitivity to properly take into account the effect on the
cross-section of the intrinsic width of the searched resonance.
Detailed Monte Carlo simulation of about 500 events per point on
the grid is performed followed by the application of the full
analysis chain.

For Monte Carlo simulation of leptoquark signals,
we make use of the LEGO event generator~\cite{LEGO} which
was described in more detail in~\cite{H1LQ94}. It
takes into account initial state QED radiation and QCD corrections
in the initial and final state and corrects properly the kinematics
at the decay vertex for effects of the parton shower masses.
The efficiencies for the leptoquark signal detection depend only
weakly on the leptoquark mass and are given in Table~\ref{tab:lqeff}
for a typical mass in the middle of the accessible range.
\begin{table}[hbt]
  \renewcommand{\doublerulesep}{0.4pt}
  \renewcommand{\arraystretch}{1.2}
  \begin{center}
    \begin{tabular}{||c||c|c|c|c||}
      \hline \hline
         & $S_{q}$ & $S_{\bar{q}}$ & $V_{q}$ & $V_{\bar{q}}$ \\
      \hline
  $LQ \rightarrow e + q$   & 46.2 & 45.5 & 48.3 & 47.3 \\
  $LQ \rightarrow \nu + q$ &      & 55.5 & 53.5 &      \\
  $LQ \rightarrow \mu + q$ & 73.0 & 71.7 & 67.3 & 56.3 \\
  $LQ \rightarrow \tau+ q \rightarrow \mu + \nu_{\mu} + \bar{\nu_{\tau}}+ q$
                           & 13.2 & 13.0 & 12.2 & 10.2 \\
  $LQ \rightarrow \tau + q \rightarrow e + \nu_e + \bar{\nu_{\tau}}+ q$
                           & 10.1 &  9.9 &  9.3 &  7.8 \\
     \hline \hline
    \end{tabular}
    \caption[~]
         {\small \label{tab:lqeff}
           Leptoquark detection efficiency in \% at $M = 150 \GeV$.}
  \end{center}
\end{table}
For leptoquarks in the $e + q$ decay channel,
the numbers include the effect of the mass dependent $y_e$ cut
and the finite width of the sliding mass bin $\Delta M_e$
which was optimized and contains about $ 68\%$ of the signal at a
given mass.
This $y_e$ cut for scalars (vectors) varies from $y_e > 0.5$
($0.25$) at $45 \GeV$ to $y_e > 0.05$ ($0.10$) at $275 \GeV$.
At $150 \GeV$, we have $y_e > 0.35$ ($0.20$) and
$\Delta M_e \simeq 25$ ($40$)$\GeV$ for scalars (vectors).
Comparable efficiencies are obtained for all leptoquarks in this decay
channel.
The efficiencies are slightly higher for leptoquarks in the $\nu + q$
decay channel where no mass dependent $y$ cut is applied.
There, at $150 \GeV$ we have $\Delta M_h \simeq 35 \GeV$ for all
leptoquarks.
In the $\mu + q$ decay channel in which we observed no events and
expected no background, the sliding mass bin could be enlarged and
contains about $ 95\%$ of the wanted signal
at a given mass while an efficiency
loss of $10\%$ comes from the ``muon track'' identification efficiency.
In the $\tau + q$ channels, the efficiencies include the leptonic
branching ratio of the $\tau$ lepton. They are slightly lower for the
$\tau \rightarrow e + \nu_e + \bar{\nu_{\tau}}$ case due to the
transverse energy requirement for the electron.

Systematic errors entering the exclusion limits derivation,
come from the uncertainty on the luminosity
%
measurement and the absolute energy calibration (see
section~\ref{sec:dislike}) and from the choice of the scale entering the
structure function calculation (leading
to $\approx 7\%$ uncertainty in the cross-section).
The choice of the parton density parametrisation implies an
uncertainty of $5\%$ at small mass and $20\%$ at largest mass on the
leptoquark cross-sections.
The effect of interference between standard DIS processes
and leptoquark boson exchange was studied and found to be negligible.

\begin{figure}[p]
  \begin{center}
    \mbox{\epsfxsize=0.9\textwidth \epsffile{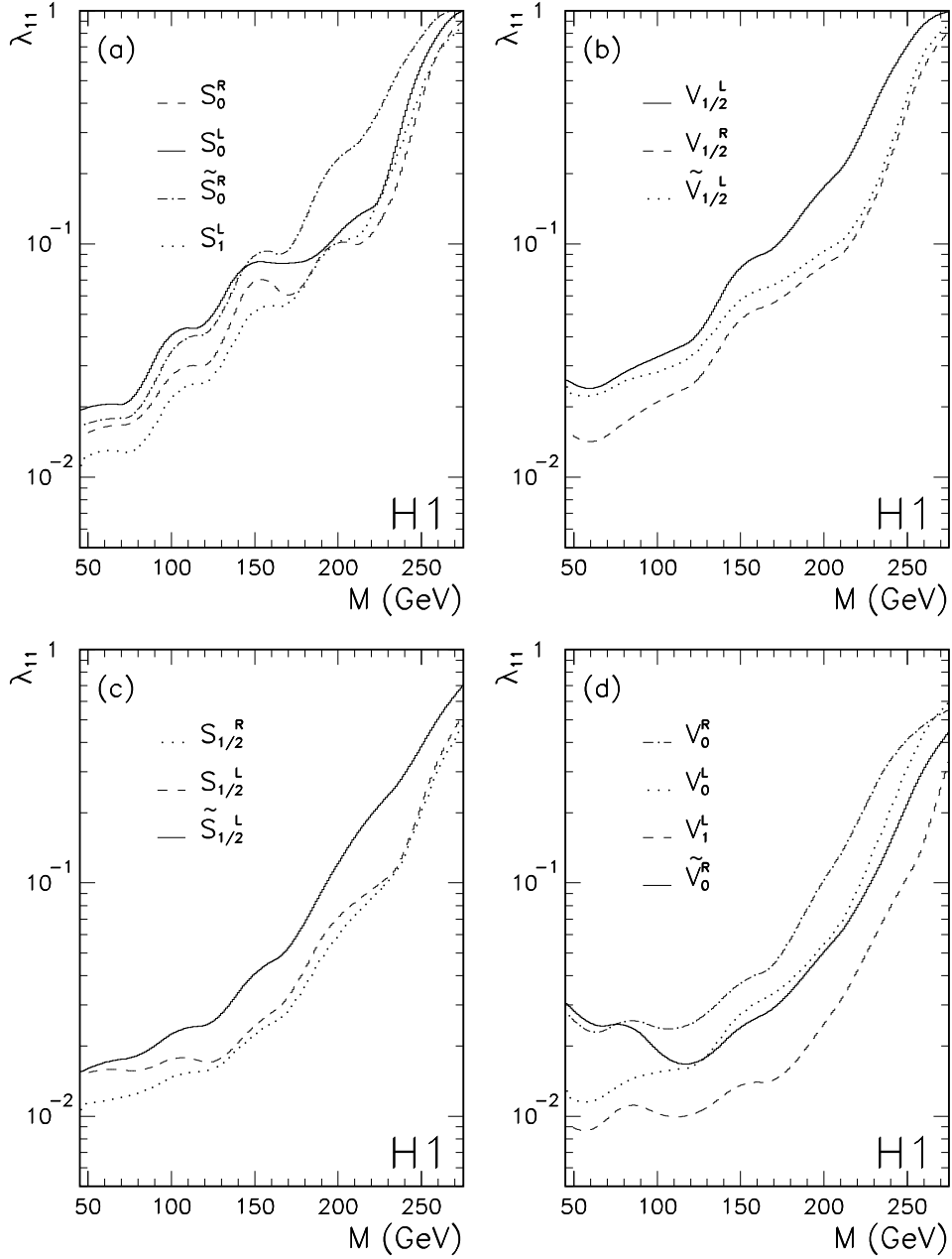}}
  \end{center}
\caption[~]
{\small   Upper limits at 95\% CL as a function of mass on the
          coupling $\lambda^{L,R}$
          for scalar and vector leptoquarks which for an $e^+$ beam
          can decay into lepton$+\bar{q}$ (a,b) and lepton$+q$ (c,d).
             The regions above the curves are excluded.
             The limits on $\lambda^L$ for $S_0, S_1, V_0$
             and $V_1$ combine $e+q$ and $\nu +q$ decays.
             Moreover, limits on $\lambda^{L,R}$ on (a) and (b)
             are obtained combining 1993 $e^- p$ and 1994 $e^+ p$ data.}
 \label{fig:lqlim94}
\end{figure}

The exclusion limits obtained on the coupling $\lambda = \lambda_{11}$
at $95\%$ confidence level (CL) are shown in Fig.~\ref{fig:lqlim94} as
a function of the leptoquark mass.
The derivation combines the two decay channels (when relevant)
with branching ratios determined from the theory~\cite{BUCH1987}.
%
%
\begin{figure}[p]
  \begin{center}
    \mbox{\epsfxsize=0.9\textwidth \epsffile{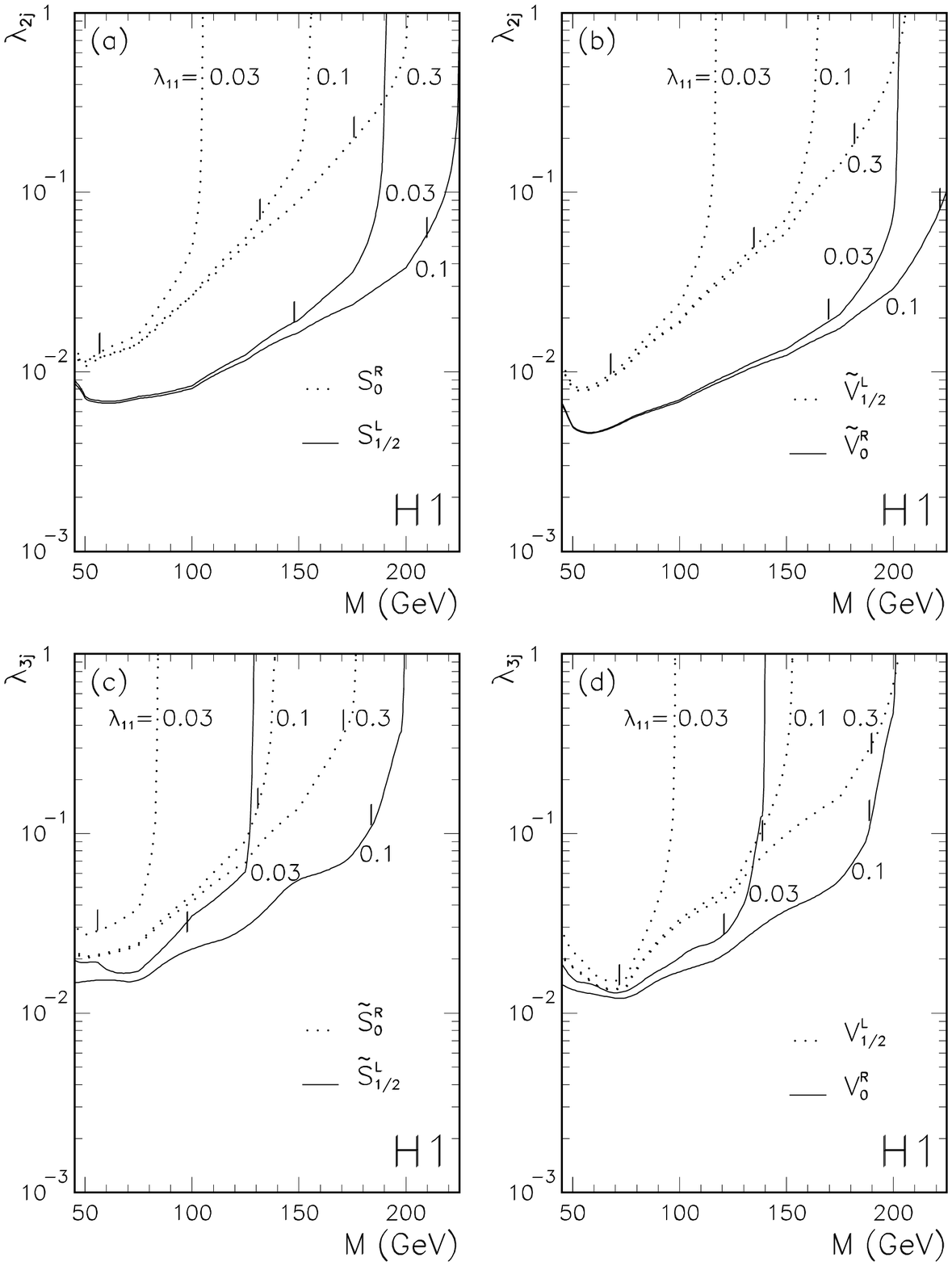}}
  \end{center}
\caption[~]
{\small   Upper limits at 95\% CL as a function of mass
          on the coupling $ \lambda_{2j}$ ($j=2,3$)
          for (a) scalars
          and (b) vector leptoquarks decaying into $\mu+q$,
          and on the coupling $ \lambda_{3j}$ ($j=1,2,3$)
          for (c) scalars
          and (d) vector leptoquarks decaying into $\tau+q$.
          The regions above the curves are excluded.
          For each leptoquark type, the limits are obtained
          for different $\lambda_{11}$ input values.
          Masses below the symbol $\mid$ are
          already excluded by the analysis in the $e+q$ channel
          for these given $\lambda_{11}$.
          The limits are obtained using 1994 $e^+ p$ data.}
   \label{fig:lq2lim}
\end{figure}
%
For leptoquarks which in the $s$-channel are produced by the fusion of
the $e^+$ with a $\bar{q}$, the exclusion domain is only slightly
improved compared to our previously published results based only on
1993 $e^-$ beam data~\cite{H1LQ94}.
In contrast, considerable improvement is obtained for other
leptoquarks which can be produced in $e^+$-{\it valence quark} fusion.
At $M = 150 \GeV$, in the middle of the accessible mass range, the exclusion
domain is extended by more than an order of magnitude.
This is partly due to the increase of integrated luminosity and
partly due to the valence versus sea quark momentum densities
in the proton.
For these most favourable cases (Fig.~\ref{fig:lqlim94}c and d),
leptoquarks with masses up to about $275 \GeV$ are excluded for
coupling strengths stronger than the electromagnetic strength
(i.e. for $\lambda_{11} \; \gsim \; 0.3 \simeq
                                        \sqrt{4 \pi \alpha_{em}}$).
Coupling values down to an order of magnitude weaker than
$\sqrt{4 \pi \alpha_{em}}$ are excluded for leptoquarks of masses
up to $200 \GeV$.

The excluded domain in the mass-coupling plane extends beyond the
mass range covered by the published results from the D0
experiment at the TEVATRON~\cite{D0}.
The limits obtained there are essentially independent of the Yukawa
couplings probed at HERA and in contrast to our results depend
significantly on the assumed branching ratios.
D0 considers only scalar leptoquarks and rejects masses below
$120$ ($133$)$\GeV$ at 95\% CL for a branching ratio of 0.5 (1.0) into a
charged lepton and a quark.

For leptoquarks possessing a coupling $\lambda_{11}$ to first generation
lepton-quark pairs as well as a coupling $\lambda_{2j}$
($\lambda_{3j}$) with leptons of the second (third) generation,
exclusion limits at $95\%$ CL are shown in Fig.~\ref{fig:lq2lim}
as a function of mass assuming different $\lambda_{11}$ input values.
Here we explicitly make use of the branching fraction into $\mu +q$
($i=2$) or $\tau + q$ ($i=3$) final states,
$ Br = \lambda_{ij}^2 / \left( \lambda_{11}^2
                             + \lambda_{ij}^2 \right) $.
The limits are plotted for representative scalar and vector
leptoquarks which couple in the $s$-channel to either valence
or sea quarks.
In the case of a non-vanishing $\lambda_{11} \lambda_{2j}$ product
(Fig.~\ref{fig:lq2lim}a and b), leptoquarks of the chosen types
are not excluded from the reach of HERA from indirect
searches~\cite{DAVIDSON}.
The decay of these four leptoquarks ($S_0^R , S_{1/2}^L ,
\tilde{V}_{1/2}^L , \tilde{V}_0^R$) via $\lambda_{23}$,
which is forbidden below the top mass, has not been considered here.
The leptoquarks chosen for  Fig.~\ref{fig:lq2lim}c and d do not
couple to the top quark.
Masses up to $225$ ($198$)$\GeV$ are excluded for the first time in a
direct search for couplings with leptons of the second (third)
generation larger than $\sqrt{ 4 \pi \alpha_{em}}$.

The published results from the CDF experiment at
the TEVATRON~\cite{CDF} exclude second generation scalar
leptoquarks with masses below $96$ ($131$)$\GeV$ at 95\% CL
with branching ratio 0.5 (1.0) into $\mu + q$ pairs.
However these limits are not directly applicable when more than one
coupling is enabled.

Finally, we make use of our knowledge of the expected background,
the signal detection efficiencies and the number of observed events
in each of the different event topologies considered in
section~\ref{sec:data7s} to derive limits on fourteen possible
combinations of two couplings.
The results are given in Table~\ref{tab:coupprod}
for a leptoquark mass of $M= 150\GeV$ and equal values of the two
couplings.
\begin{table}[htb]
  \renewcommand{\doublerulesep}{0.4pt}
  \renewcommand{\arraystretch}{1.2}
 \begin{center}
   \begin{tabular}{||rl||l|l|l|l|l|l||}
  \hline \hline
    \multicolumn{2}{||c||}{Leptoquark}
     & \Loooo & \Lootw & \Looth & \Lotot & \Lottw & \Lotth \\
 $Q$ & Type &     & $j=1,2,3$&$j=1,2,3$&      & $j=1,2,3$&$j=1,2,3$ \\
  \hline
 $ ^{-1/3}$&$S_0^L  $
               & 0.084 &  0.166   & 0.395
                                  & 0.162 & 0.319 &  0.760$_{j=2}$ \\
 $ ^{-1/3}$&$S_0^R  $
               & 0.071 & 0.118$^*_{j=2}$ & 0.162$^*_{j=2}$
                                  & 0.137 & 0.227$^*_{j=1}$
                                                 &  0.312$^*_{j=1,2}$ \\
 $ ^{-4/3}$&$\tilde{S}_0^R $
               & 0.090 & 0.107$_{j=3}$ &  0.178$_{j=2,3}$
                                  & 0.123 & 0.146 & 0.243$_{j=1,2,3}$ \\
 $ ^{-4/3,-1/3}$&$S_1^L $
               & 0.052 & 0.068$_{j=3}$ &  0.114
                                  & 0.074 & 0.098
                                                    & 0.164$_{j=2,3}$ \\
 $ ^{-5/3}$&$S_{1/2}^L $
               & 0.025 & 0.023$^*_{j=2}$ &  0.039$^*_{j=1,2}$
                                  & 0.244& 0.224$^*_{j=1}$
                                                  & 0.380$^*_{j=1,2}$ \\
 $ ^{-5/3,-2/3}$&$S_{1/2}^R $
               & 0.023 & 0.019$_{j=3}$ & 0.036$_{j=1,2,3}$
                                  & 0.140 & 0.116$_{j=3}$
                                                    & 0.219$_{j=1,2,3}$ \\
 $ ^{-2/3}$&$\tilde{S}_{1/2}^L$
               & 0.042 & 0.038$_{j=3}$ & 0.069$_{j=1,3}$
                                  & 0.157 & 0.142
                                                    &  0.257$_{j=2,3}$ \\
 $ ^{-4/3}$&$ V_{1/2}^L $
               & 0.082 & 0.076$_{j=3}$ &  0.133$_{j=1,3}$
                                  &  0.112 &  0.104$_{j=2}$
                                           & 0.182$_{j=2,3}$ \\
 $ ^{-4/3,-1/3}$&$  V_{1/2}^R $
               & 0.048 & 0.059$_{j=3}$ & 0.095$_{j=1,2,3}$
                                  & 0.074 & 0.091$_{j=2}$
                                           & 0.147$_{j=1,2,3}$ \\
 $ ^{-1/3}$&$ \tilde{V}_{1/2}^L$
               & 0.059 & 0.082$^*_{j=2}$ & 0.164$^*_{j=2}$
                                  & 0.114 &  0.158$^*_{j=1,2}$
                                                 & 0.317$^*_{j=1,2}$ \\
 $ ^{-2/3}$&$ V_0^L     $
               & 0.028 & 0.038$_{j=3}$ & 0.072$_{j=2,3}$
                                  & 0.105 & 0.142$_{j=2,3}$
                                           & 0.269$_{j=2}$ \\
 $ ^{-2/3}$&$ V_0^R     $
               & 0.038 & 0.029$_{j=3}$ & 0.051$_{j=1,2,3}$
                                  & 0.143 & 0.109$_{j=2}$
                                           & 0.192$_{j=1,2,3}$ \\
 $ ^{-5/3}$&$ \tilde{V}_0^R $
               & 0.025 & 0.017$^*_{j=2}$ & 0.031$^*_{j=1,2}$
                                  & 0.247 & 0.168$^*_{j=1,2}$
                                                   & 0.306$^*_{j=1,2}$ \\
 $ ^{-5/3,-2/3}$&$ V_1^L $
               & 0.014 & 0.012$_{j=3}$ & 0.020$_{j=1,3}$
                                  & 0.112 & 0.096$_{j=2}$
                                                   & 0.160$_{j=2,3}$ \\
  \hline \hline
  \end{tabular}
  \caption[~]
          { \label{tab:coupprod}
   {\small 95\% CL upper limits on coupling products at $M = 150 \GeV$.
    The results on diagonal couplings (\Loooo and \Lotot) are derived
    combining 1993 $e^- p$ and 1994 $e^+ p$ data. Other results rely
    on 1994 $e^+ p$ data and are valid for decays of the leptoquarks
    involving quarks of the three generations. The results marked with
    a $^*$ are only valid for the first and second quark generations.
    The quark generations $j$ for which our results are better than any
    existing limits~\cite{DAVIDSON} are indicated as subscripts in the
    table. }}
 \end{center}
\end{table}
In the case of the diagonal coupling product $\lambda_{11} \lambda_{11}$,
the limits obtained are comparable or better than any existing direct
or indirect limits~\cite{DAVIDSON,LEURER}.
In the other case with lepton flavour conserving couplings,
$\lambda_{12} \lambda_{12}$, stringent limits are established for
the first time for most of the leptoquark types.
For leptoquarks with lepton flavour violating couplings,
exclusion limits better than any existing limits~\cite{DAVIDSON} are
obtained for each leptoquark type.

\section{Conclusions}
\label{sec:conclu}

We have searched for leptoquarks with flavour conserving or
flavour violating couplings.
No evidence for the resonant production of such new particles
was found and mass dependent exclusion limits at $95\%$ confidence
level were derived for the couplings.

Two flavour conserving cases where only one sizeable coupling
exists ($\lambda_{11}$ or $\lambda_{12}$) were considered.
Leptoquarks with masses ranging from $216 \GeV$ to about $275 \GeV$
(depending on the leptoquark flavours) are excluded for
$\lambda_{11}$ coupling values larger than $\sqrt{ 4 \pi \alpha_{em}}$.
For some of the leptoquark types, the exclusion domain extends down
in coupling strength to more than an order of magnitude below
our previously published results, and far beyond the mass reach of
other existing colliders.
Stringent limits on $\lambda_{12}$ are established for the first time
for most of the leptoquark types.

Leptoquarks with lepton flavour violating couplings to the first
($\lambda_{11}$ or $\lambda_{12}$) and second ($\lambda_{2j}$) or third
($\lambda_{3j}$) generation leptons were searched for.
Masses up to $225$ ($198$)$\GeV$ are excluded for the first time in a
direct search for couplings with leptons of the second (third)
generation larger than $\sqrt{ 4 \pi \alpha_{em}}$.
Exclusion limits for combinations of two different couplings better than
any other existing limits are obtained.

\section*{Acknowledgements}
We are grateful to the HERA machine group whose outstanding efforts
made this experiment possible.
We appreciate the immense effort of the engineers and technicians who
constructed and maintain the detector.
We thank the funding agencies for their financial support of the experiment.
We wish to thank the DESY directorate for the hospitality extended
to the non-DESY members of the collaboration.

{\Large\normalsize}

\end{document}